\documentclass[twocolumn,showpacs,preprintnumbers,amsmath,amssymb,prl]{revtex4-1}
\usepackage{graphicx}
\usepackage{dcolumn}
\usepackage{bm}
\usepackage{amssymb}
\usepackage[dvips]{color}

\begin{document}
\preprint{APS/123-QED}	

\title{Drag induced lift in granular media}

\author{Yang Ding}
\author{Nick Gravish}
\author{ Daniel I. Goldman}
 \email{daniel.goldman@physics.gatech.edu}
\affiliation{%
School of Physics, Georgia Institute of Technology.\\
Atlanta, Georgia 30332, USA
}%
\date{\today}

\begin{abstract}
Laboratory experiments and numerical simulation reveal that a submerged intruder dragged horizontally at constant velocity within a granular medium experiences a lift force whose sign and magnitude depend on the intruder shape. Comparing the stress on a flat plate at varied inclination angle with the local surface stress on the intruders at regions with the same orientation demonstrates that intruder lift forces are well approximated as the sum of contributions from flat-plate elements. The plate stress is deduced from the force balance on the flowing media near the plate.

\end{abstract}

\pacs{ }
\maketitle
Objects moved through media experience drag forces opposite to the direction of motion and lift forces perpendicular to the direction of motion. The principles that govern how object shape and orientation affect these forces are well understood in fluids like air and water. These principles explain how wings enable flight through air and the fins generate thrust in water~\cite{vogel1996life}.

Lift and drag forces are also generated by movement within dry granular media, collections of discrete particles that interact through dissipative contact forces. Generation and control of these forces while moving within granular media is biologically relevant to many desert inhabitants that dive into~\cite{arnold1995ieh}, or swim within~\cite{maladen2009undulatory} sand. Lift forces are also relevant to industrial process such as soil tillage~\cite{onwualu1998draught}.


In granular media, lift and drag forces are not as well understood as in fluids; movement probes the complex fluid/solid behaviors of dense granular flows~\cite{jaeger1996granular}. While progress has been made understanding drag forces in slow horizontal and vertical drag and impact~\cite{albert99, *golAumb08, *nelson2008projectile, *gravishVF2010}, there has been comparatively little work investigating lift forces. Studies have examined lift forces for a partially submerged vertical rod moving horizontally and for a rotating plate~\cite{soller2006drag, wieghardt1974forces}, and the drag force on submerged objects with curved surfaces~\cite{albert2001gdd}; however, the lift forces experienced by horizontally translated submerged intruders have not been explored.

\begin{figure} [htp]
\includegraphics[width=0.4\textwidth]{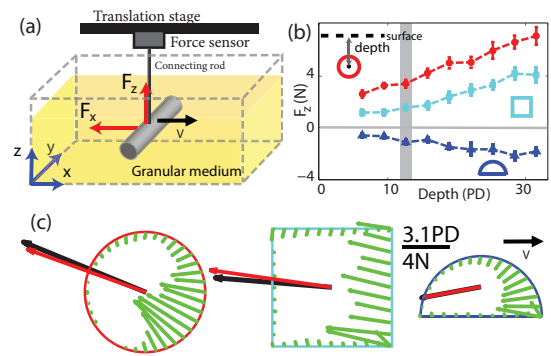}
\caption{(color online) Lift and drag forces in granular media: (a) Schematic of the experimental setup. (b) Lift force as a function of depth for the cylinder ({\color{red} $\bullet$}), square rod ({\color{cyan} $\blacksquare$}) and half cylinder ({\color{blue} $\blacktriangle$}). Gray region indicates the depth at which forces in (c) were measured. (c) Net force on rods measured in experiment ({\color{black}$\leftarrow$}) and simulation ({\color{red}$\leftarrow$}). Forces ({\color{green}$\leftarrow$}) on the intruder surfaces were measured in simulation and are scaled by four for better visibility.}
\end{figure}

{\em Experiment and simulation--} Experiment and simulation were employed to investigate the lift ($F_z$) and drag ($F_x$) forces on simple shapes during horizontal translation in granular media (Fig.~1). In experiment long intruders with different cross-sections were dragged within a bed of glass beads with particle diameter (PD) of $0.32 \pm 0.02$~cm and density ($\rho$) $2.47$~g/cm$^3$ (Fig.~1). Dragging was performed at a constant speed $10$~cm/sec with the intruder's vertical mid-point at depth $d=12.5$~PD and its long axis perpendicular to the direction of motion. In experiment, $l=31.3$~PD long intruders were connected at the midpoint to a force sensor (mounted to a linear translation stage) by a stiff stainless steel rod of diameter $2$~PD. Following the method of~\cite{albert99}, forces on the connecting rod were determined in separate measurements and subtracted from $F_x$ and $F_z$. The grain bed was 75~PD wide by 53~PD deep by 75~PD long. The initial packing state of the grains was prepared by shaking the container moderately in the horizontal direction before each run. The volume fraction was determined through measurements $\rho$, total grain mass (M), and occupied volume (V) to be $\frac{M}{\rho V} = 0.62\pm0.01$ .

Simulation employed the soft-sphere Discrete Element Method (DEM)~\cite{lee1993angle} method in which particle-particle and particle-intruder contact interactions were determined by the normal force $F_{n}=k\delta^{3/2}-G_{n}v_{n}\delta^{1/2}$ and the tangential force $F_{s}=\mu F_{n}$, where $\delta$ is the ``virtual overlap'' and $v_n$ is the normal component of the relative velocity. $F_{n}$ comprises a Hertzian contact term and a velocity dependent normal dissipation \cite{lee1993angle}. Constants $k=2\times10^6$~kg\,s$^{-2}$m$^{-1/2}$, $G_n=15~$kg\,s$^{-1}$m$^{-1/2}$ and $\mu=\mu_{\{pp,pi\}}=\{0.1,~0.27\}$ represent the hardness, viscoelastic constant, and particle-particle and particle-intruder friction coefficients respectively~\footnote[1]{To speed up computation a smaller $k$ was used than that of glass beads ($k_{glass} \approx 2\times$10$^9$~kg\,s$^{-2}$m$^{-1/2} $). $k$ was chosen so that $\delta$ is less than 0.2\%~PD for all conditions. Doubling $k$ changed net forces by less than 5\%. $G_n$, $\mu_{pp}$, and $\mu_{pi}$ were chosen to match experimental measurements of restitution coefficient ($0.92 \pm 0.05$ at 40~cm/s), and friction coefficients.}. The simulated grain bed, bounded by frictionless walls, was 75~PD wide by 55~PD deep by 78~PD and consisted of 350,000 particles in a bi-disperse mixture of equal parts 0.3 and 0.34~cm diameter spheres. Doubling any dimension of the grain bed did not change the forces significantly. The initial volume fraction $0.62\pm0.01$ was prepared by randomly distributing the particles in the volume corresponding to the desired volume fraction and then eliminating particle overlap~\cite{o2002random}. Using lower volume fractions reduced the magnitude of the forces but gave qualitatively similar results. Forces were nearly independent of speed, like other drag studies in the non-inertial regime~\cite{wieghardt1975experiments,chehata2003dgf} (speed $\lessapprox40$~cm/s for our study). In simulation the intruder dimensions and intruder speed were matched to those in experiment and the forces were averaged over the steady-state time interval.

{\em Shape determines lift--} In both experiment and simulation, $F_z$ was sensitive to the cross-section of the intruder. As shown in Fig.~1(b,c), $F_z$ for the half-cylinder was downward (opposite the orientation of the curved surface), while for the two vertically symmetric geometries, the full cylinder and square rod, $F_z$ was positive with magnitude larger for the cylinder than for the square. Experiments with smaller glass beads (0.3mm) gave similar results (see supplementary material). $|F_z|$ increased with intruder depth for all intruders (Fig.~1b). The lift mechanism is different from the Brazil nut effect~\cite{rosato1987brazil} which results from agitation of the medium by the container.

Simulation allowed investigation of the surface stress distribution responsible for lift and drag on the intruders. For all shapes the surface stress was largest along the leading surface (Fig.~1c green arrows). Due to the linear dependence of granular pressure with depth and the finite size of the intruder~\cite{albert99}, local stress increased with depth along the flat face of the square (Fig.~1c). However, for curved intruders (e.g. the cylinder in Fig.1c), the magnitude of local stress was primarily determined by the local surface orientation. As the local surface tangent became more aligned with the intruder velocity the force magnitude became small, supporting observations that surfaces parallel to the direction of motion contribute little to the drag force~\cite{albert2001gdd}. Since the normal force was larger than the frictional force, the direction of the local grain-intruder reaction force was nearly opposite to the surface normal at all points along the intruder's leading surface.

The dependence of the forces on the local surface orientation suggests that decomposing the surfaces into differential area elements and summing the forces on those elements may describe the net drag and lift experienced by the three shapes studied. A similar decomposition was successfully used to calculate net drag and thrust on an undulatory sand-swimmer~\cite{maladen2009undulatory} in the horizontal plane.

\begin{figure} [thpb]
\includegraphics[width=0.35\textwidth]{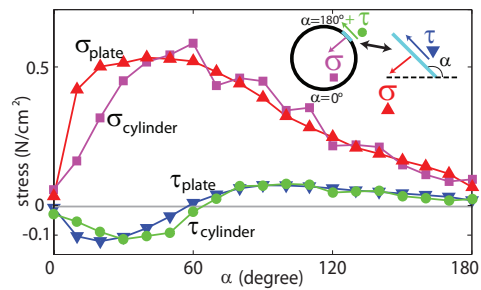}
\caption{Normal ($\sigma$), and shear ($\tau$), stress on the leading surface of the cylinder as a function of tangent angle $\alpha$ compared to the stresses on a plate with the same $\alpha$. {\color{magenta} $\blacksquare$}, {\color{red}$\blacktriangle$}, {\color{green}$\bullet$}, and {\color{blue}$\blacktriangledown$} represent $\sigma_{cylinder}$, $\sigma_{plate}$, $\tau_{cylinder}$ and $\tau_{plate}$ respectively.}
\end{figure}

{\em Plates as differential elements--} To determine if the forces on curved intruders can be understood by decomposing the shape into flat plate elements we now study the stresses on a flat plate with tangent angle $\alpha$ varied between $0^\circ$ and $180^\circ$ (e.g. $\alpha=0^\circ$ is along the direction of motion). In simulation, a long ($l=31.3$~PD), thin ($0.1$~PD), flat plate of finite width $w=7.94$~PD was dragged horizontally through the granular medium with its center 12.5~PD below the initial surface, and the average normal ($\sigma$) and shear stress ($\tau$) on the leading side of the plate were measured as a function of $\alpha$ (Fig.~2). The stresses were asymmetric about $\alpha=90^\circ$, and $\sigma$ increased rapidly for small $\alpha$, peaking at $\alpha\approx50^\circ$. At $\alpha \approx 60^\circ$, $\tau$ changed sign, indicating a reversal in grain flow along the surface. We define the effective friction ratio on the plate as $\mu_{peff}(\alpha)=\tau/\sigma$, which is zero at $\alpha\approx 60 ^\circ$ and saturates to the expected magnitude of $\mu_{pi}$ for $\alpha>135^\circ$ and $\alpha\approx0^\circ$ with opposite signs. Remarkably, along the surface of the cylinder (half-cylinder and square rod as well), the stresses approximately matched the stresses on the plate oriented at the same angle $\alpha$ (Fig.~2). The stresses on the intruders were corrected by considering that the depth of the differential element is different from the depth of the center of the cylinder and assuming linear dependence of the stress on depth.

\begin{figure} [htpb]
\includegraphics[width=0.4\textwidth]{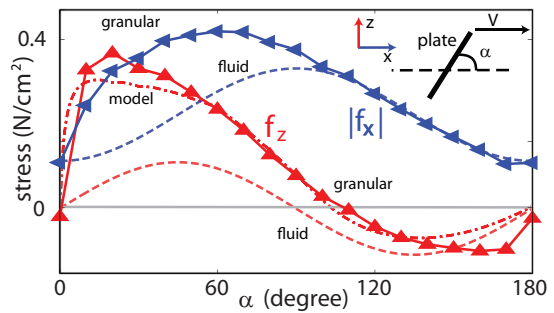}
\caption{(color online) (a) The drag ($|f_x|$, blue) and lift ($f_z$, red) components of the stress on a plate as a function of $\alpha$ in granular media ({\color{blue}$\blacktriangleleft$} and {\color{red}$\blacktriangle$}) as compared to a fluid with with Re~$\ll1$ (dashed lines)~\cite{gombosi1994gaskinetic}. Dash-dot red line is granular wedge model (see {\em Force model} section).}
\end{figure}

The near equality between the stresses on plates and local surface regions of intruders with the same orientation implies that $F_z$ and $F_x$ for a translated rod can be approximated by the sum of forces from the corresponding shape built from infinitesimal plates. Resolving the stresses on the plate at orientation $\alpha$ into the lab frame ($xyz$) gives the local lift $f_z(\alpha)$ and local drag $f_x(\alpha)$ force contributions per unit area on the plate element (Fig.~3). $F_z$ for different shaped rods results from the integration of $f_z(\alpha)$ contributions along the intruders leading surface, corrected by a linear depth term, over the intruders infinitesimal surface area $dA$, e.g. $F_z = \int f_z(\alpha)(z/d)dA$ (Fig.~4a). Comparison of this integration over the three rod shapes with the measured $F_z$ from simulation and experiment (Fig.~4b) shows good agreement. The orientation of the leading surface of the cylindrical intruder varies from $0\leq \alpha \leq 180^\circ$ and the asymmetry of $f_z(\alpha)$ results in a net positive lift. Positive $f_z$ at $\alpha=90^\circ$ is responsible for the small lift on the square rod. For the half-cylinder, $90\leq \alpha \leq 180^\circ$, and $F_z$ is negative.

\begin{figure} [htpb]
\includegraphics[width=0.47\textwidth]{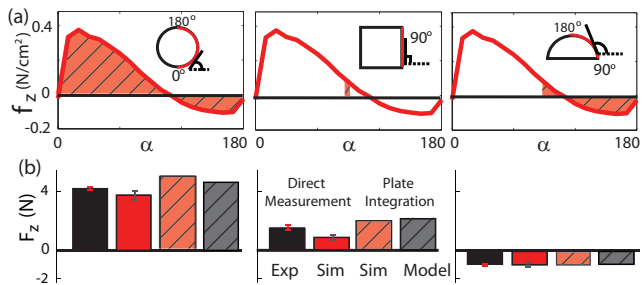}
\caption{(color online) (a) $F_z$ on intruders calculated by integration of $f_z(\alpha)$. Red outlines on the shapes indicate the leading surface and the hatched area indicates the corresponding range of $\alpha$ and region of integration. (b) $F_z$ calculated by integration of $f_z$ from simulation (red hatched bar) and model (gray hatched bar) compared to direct measurement of $F_z$ in experiment (black bar) and simulation (red bar).}
\end{figure}

To gain insight into the nature of granular drag and lift we compare our results to those from low Reynolds number fluids where inertia is negligible and viscous forces dominate. $f_x$ and $f_z$ on a plate at angle $\alpha$ in low Re fluids are symmetric along the direction of motion $\alpha=90^\circ$ (Fig.~3, dashed curves) while the drag and lift forces on the plate in granular media are asymmetric about $\alpha=90^\circ$. This suggests that a granular model is required to understand the origin of lift in granular media.

\begin{figure} [htp]
\includegraphics[width=0.45\textwidth]{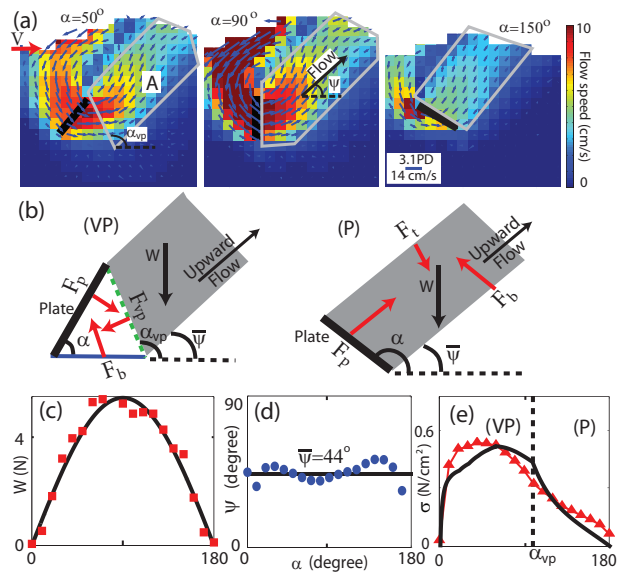}
\caption{(color online) Flow of grains and force-balance model: (a) Flow field in the vertical ($xz$) plane for three plate (solid black line) orientations. Gray boundary indicates regions with upward flow~[18]. (b) Forces on a wedge for $\alpha<\alpha_{vp}$ [regime (VP)]  and on the band for $\alpha\geq \alpha_{vp}$ [regime (P)]. (c) The weight of the upward-flow region as function of $\alpha$ calculated from simulation ({\color{red} $\blacksquare$}) and fit (black) to $W=c\sin(\alpha)$, where $c=5.7$~N. (d) The average flow velocity angle $\psi$ in the upward-flow region vs. $\alpha$ ({\color{blue}$\bullet$}).  (e) Normal component of the stress on the plate $\sigma$ calculated from the model (black) and measured from simulation ({\color{red}$\blacktriangle$}). The black dashed line $\alpha=\alpha_{vp}=97^\circ$ indicates the boundary between the two regimes (VP) and (P).}
\end{figure}

{\em Force model--}  In the quasi-static regime of granular flow, the force on an intruder can be determined by analyzing the force balance on the moving volume of grains~\cite{wieghardt1975experiments,onwualu1998draught}. With the plate acting as one flow boundary we can determine the normal and tangential components of stress ($\sigma$ and $\tau$) on the plate surface from these force balance equations. In practice this requires approximating the boundaries of the moving media as planes and computing the forces acting on them.

Examination of the motion of grains in the vertical plane ($xz$) reveals that the particles move upwards in front of the plate and flow along a lower boundary (see Fig.~4a), a slip plane. Finite yield stress in granular media results in flowing regions bounded by slip planes with upward flow direction due to increasing yield stress with depth~\cite{wieghardt1975experiments}. The upper region of the flow is confined by a boundary starting from the top edge of the plate and approximately parallel to the lower boundary. The weight $W$ of this up-flow region $A$ \footnote[2]{\label{defineflow}The upwards flow area $A$ is defined as the region where average particle velocity $\vec{v}$ satisfies $v_x>0.1V$ and $v_z>0$. Velocity is averaged over 1.9~PD$\times$1.9~PD cells along the central 90\% length of the plate at four time instants with intruder position separated by 0.3~PD. $W$ is calculated as $W=\rho_{e} g l A$, where $\rho_{e}=1.53~$g/cm$^3$ is the effective media density including voids.} is well fit by $W = c\sin\alpha$, and is thus proportional to the projected plate length normal to the direction of motion (see Fig.~5c). The direction of average velocity of the particles in the band $\psi$ varies little (see Fig.~5d) and we approximate the angle of the flow boundaries as $\bar{\psi}=44^\circ$.

The plate defines one of the boundaries of $A$ for large $\alpha$ (e.g. $\alpha=150^\circ$ in Fig.~5a) but for small $\alpha$ (e.g. $\alpha=50^\circ$ in Fig.~5a), the vertical velocity of the particles adjacent to the plate is negative and thus the upward flow boundary is described by a virtual plane intersecting the top of the plate and extending downwards at angle $\alpha_{vp}$. We therefore consider two regimes of flow: for $\alpha\ge\alpha_{vp}$, we approximate $A$ as bounded by the plate and two parallel surfaces with angle $\bar{\psi}$ [regime (P) in Fig.~5b]. For $\alpha<\alpha_{vp}$, the upwards flow region $A$ is bounded by the virtual plane and two parallel surfaces with angle $\bar{\psi}$. The region adjacent to the plate is bounded by the plate, the virtual plane, and an approximately horizontal bottom surface [regime (VP) in Fig.~5b]. For regime (P), the forces on the flowing band are the forces from the top and bottom boundaries and its weight. On the plate surface, the friction coefficient $\mu_{peff}(\alpha)$ is used. Within the media, dynamic friction is assumed and the friction coefficient is $\tan\gamma$, determined by the angle of repose, $\gamma=13\pm1^\circ$\footnote[3]{This value is experimentally $\gamma=20\pm2^\circ$, we attribute the difference to the perfect spherical shape of the grains in simulation.}, which is measured in separate simulations by tilting the initially horizontal container and recording the post avalanche surface orientation. Simulation indicates that the stress at the bottom surface dominates; therefore we neglect the force on the top of the flow band ($F_t$ in Fig.~5b). The normal stress on the plate, $\sigma$, can be solved from the force balance on the band to obtain $\sigma(\alpha)=\frac{W}{lw}\frac{\cos\beta\sin(\bar{\psi}+\gamma)}{\sin(\alpha-\beta-\bar{\psi}-\gamma)}$, where $\beta(\alpha)\equiv\tan^{-1}\mu_{peff}$.

For regime (VP), the normal stress on the plate is calculated by considering the stress on the wedge adjacent to the plate. The stress on the virtual plane can be solved by the above equation with the corresponding weight of the band and a fixed angle $\alpha=\alpha_{vp}$. Solving the force balance equation for this triangular wedge we obtain $\sigma(\alpha)=\frac{W}{wl}\frac{\cos\beta\sin(\bar{\psi}+\gamma)}{\sin(\alpha_{vp}-\beta'-\bar{\psi}-\gamma)}\frac{\sin(\alpha_{vp}-\beta'-\gamma)}{\sin(\alpha-\beta-\gamma)}$, where $\beta'=\beta(\alpha_{vp})$. The parameter $\alpha_{vp}=97^\circ$, determined from a best fit of $\sigma(\alpha)$ from simulation, is within the expected range from flow field observations (Fig.~5a).

Comparison of $\sigma$ calculated from the wedge model and $\sigma$ measured directly in simulation demonstrates that the model captures the asymmetric shape of $\sigma$ (Fig.~5e). Integration of $f_x$ calculated from the model over the non-planer intruder surfaces yields net lift forces in agreement with those from other methods (Fig.~4b). The decrease in $\sigma$ above $\alpha=90^\circ$ results from the decrease in $W$ with increasing $\alpha$. For $\alpha \leq 90^\circ$, although $W$ increases with $\alpha$, the stress on $A$ is transmitted by the wedge which induces extra resistance on the bottom plane; therefore $\sigma$ peaks at $\alpha$ smaller than $90^\circ$. The model assumes the up-flow area $A$ ($W$) increases with depth which explains the monotonic increase of $|F_z|$ with depth. The discrepancy between $\sigma$ in model and simulation may be due to the simplified description of the shape of the boundaries of the flowing media, approximation of $\psi$ and $\alpha_{vp}$ as constants and $W$ as a simple function, and neglecting $F_t$.

Decomposition of the force on the intruder into forces on differential elements assumes that the flowing region corresponding to each element is not disturbed by other elements. This may explain the difference we observe between the stress on the plate and the local stress along the intruder for small $\alpha$ (corresponding to elements on the bottom), where the upper flow boundary for these elements is obstructed by higher regions of the cylinder. For objects with concave leading surfaces, the use of differential surface elements may require consideration of material jamming in the concave region.

We have shown that the magnitude and sign of the drag induced lift force in granular media depends on the shape and depth of the intruder. Drag induced lift on non-planar intruders can be computed as the summation of lift forces from planar elements which each experience a lift force resulting from the pushing of material up a granular slip plane. The increase of yield stress with depth in granular media causes the asymmetric flow and enhances the lift force on a plate facing downward. Our understanding of these forces can elucidate the effects of head and body shapes~\cite{mosauer1932adaptive} of sand burrowing organisms, and could aid design of control surfaces to allow robots~\cite{synthsandswim} to maneuver in granular environments.


We thank Ryan D. Maladen, Paul Umbanhowar and Chen Li for discussion. This work was supported by NSF Physics of Living Systems grant PHY-0749991, the ARL MAST CTA and the Burroughs Wellcome Fund.
\bibliographystyle{apsrev4-1}
%
\end{document}